 \documentclass[twocolumn]{aastex631}
 \usepackage{booktabs}
 \usepackage{natbib}
 \usepackage{booktabs}
 \usepackage{amsmath}
 \usepackage{hyperref}
 \usepackage{cleveref}
\usepackage{amsmath,graphicx}
\usepackage{mathrsfs}

\begin{document}


\title{Possible  Multi-band Afterglows of FRB 20171020A and its Implication }
\author{Ke Bian}
\affiliation{Guangxi Key Laboratory for Relativistic Astrophysics, Department of Physics, Guangxi University, Nanning 530004, China;dengcm@gxu.edu.cn}
\author{Can-Min Deng}
\affiliation{Guangxi Key Laboratory for Relativistic Astrophysics, Department of Physics, Guangxi University, Nanning 530004, China;dengcm@gxu.edu.cn}


\begin{abstract}
	Fast Radio Bursts (FRBs) are millisecond-duration radio transients of mysterious origin, with growing evidence linking at least some of them to magnetars. While FRBs are primarily observed in the radio band, their potential multi-wavelength afterglows remain largely unexplored. We investigate the possible afterglow of FRB 20171020A, a rare nearby and bright FRB localized in a galaxy at only 37 Mpc. Assuming that this source produces a future bright burst, we model the expected afterglow emission in the radio, optical, and X-ray bands under both uniform and wind-like ambient media, within the framework of the magnetar model. Our results show that the optical afterglow is the most promising for detection, but it fades rapidly and requires follow-up within a few hundred seconds post-burst. The radio afterglow may be detectable under favorable conditions in a dense stellar wind, whereas the X-ray counterpart is too faint for current telescopes. These findings suggest that rapid optical follow-up offers the best opportunity to detect the afterglow of the next bright burst from FRB 20171020A, providing unique insights into the progenitor and its environment. To assess observational feasibility, we estimate the event rate of nearby FRBs with sufficient energy to power detectable afterglows, finding a rate of $\sim$0.3 per year for CHIME surveys. Although this rate is low and the optical detection timescale is short, coordinated fast-response strategies using global telescope networks could significantly improve the chance of success.  As more nearby FRBs are discovered, multi-wavelength observations will be essential in unveiling the physical nature of these enigmatic events.
\end{abstract}



\section{Introduction}

{Fast Radio Bursts (FRBs) are millisecond-duration radio transients of extragalactic origin, exhibiting extreme luminosities and diverse observational properties. Since their discovery \citep{Lorimer:2007qn,2013Sci...341...53T}, extensive efforts have been made to understand their progenitors and emission mechanisms \citep{2021SCPMA..6449501X,2023RvMP...95c5005Z,2024Ap&SS.369...59L}. While magnetars have emerged as a leading candidate for at least some FRBs, as~evidenced by the association of FRB 200428D with the Galactic magnetar SGR 1935+2154 \citep{2020Natur.587...54C,2020Natur.587...59B,Lin_2020}, the~observed diversity in FRB properties—including repetition rates, polarization variability, and~host galaxy environments—suggests multiple progenitor channels or environmental influences \citep{2019Natur.566..235C,2019ApJ...887L..30K,2021Natur.598..267L,2022Natur.602..585K,2021ApJS..257...59C,2021ApJ...923....1P,2021NatAs...5..594N,doi:10.1126/science.abl7759,2022Natur.609..685X,2023ApJ...947...83C}.
}

{Recent large-sample surveys, such as those conducted by CHIME/FRB, have greatly expanded our knowledge of FRBs, revealing a wide range of behaviors and host galaxy types. Some FRBs are found in dwarf star-forming galaxies, while others are located in globular clusters or older stellar populations. This diversity suggests that the FRB population likely consists of multiple progenitor classes \citep{2023RvMP...95c5005Z}. Proposed progenitor models span magnetar flares, neutron star mergers, accretion-induced collapses, white dwarf interactions, and~even exotic scenarios such as primordial black hole evaporation \citep{Platts2020,Zhang2020}. A~growing number of FRBs are found to repeat, while others show only a single detected burst, further complicating the picture.
}

{One of the promising approaches to unraveling the nature of FRBs is the search for their multi-wavelength afterglows \citep{2014ApJ...792L..21Y,2021Univ....7...76N}. Similar to gamma-ray bursts, FRB explosions may eject relativistic material, which interacts with the surrounding medium to produce afterglow emission across the electromagnetic spectrum \citep{Metzger+2019,Beloborodov+2020,2020ApJ...899L..27M}. The~detectability and evolution of such afterglows depend strongly on the energy of the ejected material and the properties of the circum-burst environment.
	Recent studies have shown that afterglow observations could help distinguish between different FRB models. For~instance, \citet{2022MNRAS.517.5483C} examined the lack of a detected afterglow from FRB 200428D and argued that a simple, constant-density environment was unlikely, while a wind-like medium could better explain the observations. \citet{2023RAA....23k5010D} also conducted an analysis of the multi-band afterglow of FRB 200428D using the standard external shock synchrotron emission model commonly applied to gamma-ray bursts.
}

{Despite growing theoretical interest, afterglow detections remain elusive, likely due to the short duration and low energy output of typical FRBs, as~well as the often large distances involved. Detectability may be improved for nearby and unusually bright FRBs, which provide a rare opportunity to probe the surrounding medium and the nature of the outflows. In~this context, nearby FRBs with high fluence are of particular \mbox{observational interest}.}

{Here, we focus on FRB 20171020A, an~exceptional FRB discovered by ASKAP \citep{2018ApJ...867L..12S} and localized in its host galaxy, ESO 601-G036, at~a distance of just 37 Mpc \citep{2018ApJ...867L..10M,2023PASA...40...29L}. This makes it {one of the closest extragalactic FRBs.}  What makes FRB 20171020A even more remarkable is its extraordinary brightness of 200 Jy ms, ranking it among the brightest FRBs localized by ASKAP so far \citep{2025PASA...42....3A}. Its unique combination of proximity and high brightness makes it a prime candidate for afterglow studies, presenting a rare opportunity to probe the circum-burst environment and relativistic ejecta energetics, given its high fluence suggesting a substantial energy reservoir for detectable afterglows.}

{In this work, we conduct a theoretical investigation into the potential multi-wavelength afterglow of FRB 20171020A. By~modeling the dynamics of the FRB ejecta and its interaction with different external environments—specifically, a~uniform interstellar medium and a stellar wind—we derive the expected light curves in the radio, optical, and~X-ray bands and compare them with the sensitivity limits of current and upcoming observational facilities. {It is important to emphasize that no afterglow was detected following the previous burst of FRB 20171020A due to the lack of timely follow-up observations. Therefore, our predictions pertain to the afterglow that might accompany its next burst in the future.} Given the source’s relatively close distance and high fluence, FRB 20171020A stands as an excellent candidate for targeted follow-up observations. This study not only deepens our understanding of FRB 20171020A but also provides a framework for optimizing follow-up strategies for nearby, bright FRBs, which in turn can help constrain FRB progenitor models and the characteristics of their surrounding environments.
}

\section{Model~Description}\label{sec2}

To investigate the possible multi-wavelength afterglow of FRB 20171020A, we adopt the flaring magnetar model, in~which a relativistic outflow from a magnetar collides with the surrounding medium, producing synchrotron maser emission at the shock front \citep{Metzger+2019,2020ApJ...899L..27M}. This process not only generates the prompt FRB emission, but~also deposits energy into the external medium, potentially leading to a long-lived afterglow. Following this framework, we model the interaction between the FRB ejecta and its environment, and~calculate the resulting afterglow light curves in the radio, optical, and~X-ray~bands.  

\subsection{Dynamics of the~Ejecta}
Following the production of FRB emission, the~relativistic ejecta continues to propagate outward, sweeping up ambient material and producing synchrotron afterglow emission. We consider this relativistic ejecta with an initial kinetic energy $E_k$ and an initial Lorentz factor $\eta$.  As~the ejecta propagates into the surrounding medium, it undergoes deceleration due to the accumulation of swept-up material. 
In order to unify the description of the early and late dynamical evolution of the blast wave, we adopt the formula given \mbox{by \citet{10.1046/j.1365-8711.1999.02887.x}}
\begin{equation}
\frac{\mathrm{d}\Gamma}{\mathrm{d}M_{\rm s}}=-\frac{\Gamma^2-1}{M_{\mathrm{ej}}+\epsilon M_{\rm s}+2(1-\epsilon)\Gamma M_{\rm s}},
\end{equation}
where $\Gamma$ is the  Lorentz factor of the blast wave, and $\epsilon$ is the fraction of the shock-generated thermal energy that is radiated.
\( M_{\rm s} \) represents the mass inside a given radius \(r\), which is defined by
\begin{eqnarray}
M_{\rm s} = \int_{0}^{r} m_{\rm p}n(r) 4\pi r^2 \, {dr}' ,
\end{eqnarray}
where the number density  \(n(r)\) of the surrounding medium scaling as $	n(r) \propto r^{-k}$. 	$k=0$ represents the  uniform medium, and~$ k=2$ represents the wind-like medium.  One can derive the relationship between \(\Gamma\) and \(r\) by substituting $M_{\rm s}$ into Equation~(1).
Moreover, using the relation of $r = {ct}/({1 - \beta})$,
one can derive the relationship between \(\Gamma\) and time \(t\) of the observer, where $\beta$ is the velocity of the shocked  material, and~$\Gamma=(1-\beta^2)^{-1/2}$.

\subsection{Thermal Synchrotron~Emission}
As the ejecta decelerate, a~forward shock {forms} 
{In the parameter space considered in this work, the~lifetime of the reverse shock is very short, in~the order of  ms. 
	Given this short duration, we focus solely on the afterglow emission produced by the forward shock.}, 
accelerating electrons and amplifying magnetic fields. These relativistic electrons then produce synchrotron radiation across multiple wavelengths, resulting in a broadband afterglow.
In the standard afterglow models of gamma-ray bursts, relativistic shocks are often assumed to accelerate electrons into a non-thermal power-law distribution via the diffusive shock acceleration process. However, in~the case considered in this work, the~external medium surrounding FRBs  is expected to be highly magnetized \mbox{($\sigma \gtrsim 0.1$) \citep{Metzger+2019,2020ApJ...899L..27M}}. In~such an environment, the~strong magnetic field suppresses particle diffusion across the shock front, leading to the breakdown of the standard Fermi acceleration mechanism \citep{2009ApJ...698.1523S}. As~a result, instead of a non-thermal power-law distribution, the~electrons in the shocked region are expected to follow a thermal distribution characterized by
\begin{eqnarray}
\gamma_{\rm s} = \frac{m_{\rm p}}{2m_{\rm e}} \Gamma ,
\end{eqnarray}
where \(m_{\rm p}\) and \(m_{\rm e}\) are the proton and electron masses, respectively. 
In a magnetized shock, the~downstream magnetic field {$B$} is amplified due to compression and can be estimated as
\begin{eqnarray}
B=\sqrt{32 \pi \sigma  \Gamma^2 n m_p c^2},
\end{eqnarray}
where $\sigma$  represents the magnetization  of the upstream medium{.} {We adopted} $\sigma = 10^{-1}$ in this  work following \citep{2022MNRAS.517.5483C}, as~required for coherent maser emission for FRBs \citep{Metzger+2019}.

For synchrotron radiation of an electron with Lorentz factor \(\gamma_e\), the~observed radiation power and the characteristic frequency  are given by
\begin{eqnarray}
P(\gamma_e) &\simeq \frac{4}{3} \sigma_T c \Gamma^2 \gamma_e^2 \frac{B^2}{8\pi},
\end{eqnarray}
and
\begin{eqnarray}
\nu(\gamma_e) &\simeq \Gamma \gamma_e^2 \frac{q_e B}{2\pi m_e c},
\end{eqnarray}
where \(q_e\) is the electron~charge.

For an individual electron, the~spectral power \(P_\nu\) (in erg Hz\(^{-1}\) s\(^{-1}\)) varies as \(\nu^{1/3}\) for \(\nu < \nu(\gamma_e)\), and~cuts off exponentially for \(\nu > \nu(\gamma_e)\) \citep{1976PhFl...19.1130B}. The~peak power occurs at \(\nu(\gamma_e)\), with~an approximate value of
\begin{eqnarray}
P_{\nu,\max} \approx \frac{P(\gamma_e)}{\nu(\gamma_e)} = \frac{m_e c^2 \sigma_T}{3 q_e} \Gamma B.
\end{eqnarray}

The synchrotron radiation spectrum  is characterized by a multi-segment broken power law with an exponential cutoff at higher frequencies, with~key frequencies: the critical synchrotron frequency \(\nu_s\), the~cooling frequency \(\nu_c\), and~the self-absorption frequency \(\nu_a\). 

The cooling frequency is given by
\begin{eqnarray}
\nu_{\rm c} &=& \frac{ q_e B}{ m_e c}\Gamma \gamma_c^2.
\end{eqnarray}

The critical synchrotron frequency of the shocked electrons is
\begin{eqnarray}
\nu_{\rm s} &=& \frac{ q_e B}{2\pi m_e c}\Gamma \gamma_s^2.
\end{eqnarray}

The self-absorption frequency is given by
\begin{eqnarray}
\nu_a = \left\{
\begin{array}{ll}
\left[ \frac{c_1 (p - 1)}{3 - k} \frac{q_e n R}{B \gamma_s^5} \right]^{3/5} \nu_s, & \text{if } \nu_a < \nu_s, \\
\left[ \frac{c_2 (p - 1)}{3 - k} \frac{q_e n R}{B \gamma_s^5} \right]^{2/(p + 5)} \nu_s, & \text{if } \nu_s < \nu_a < \nu_c.
\end{array}
\right.
\end{eqnarray}
where $c_1,c_2$ are coefficients that depend on $p$ \citep{2003MNRAS.342.1131W}.

The luminosity observed by the observer is given by
\begin{eqnarray}
L_{\nu,\max} &=& N_e P_{\nu,\max},
\end{eqnarray}
where $N_e={M_s}/{m_p}$ is the total number of electrons participating in the~radiation.

Since the electrons remain thermalized rather than forming a power-law tail, their synchrotron emission is expected to differ significantly from conventional afterglow models, with~a spectrum more closely resembling thermal synchrotron radiation rather than the usual broken power-law shape.
In the context of this study, the~electrons undergo fast cooling  during the early stage, which switches to slow cooling at the later stage. The~timescale at which this transition occurs is on the order of tens of seconds.
At early times, the~electrons are in the fast-cooling regime ($\nu_{\rm c} < \nu_{\rm s}$), where~the spectral flux {is} \mbox{given by \citep{Sari_1998,GAO2013141}}\vspace{6pt}
\begin{eqnarray}
\begin{array}{l}
L_{\nu}=\left\{\begin{array}{ll}
\left(\nu/\nu_{\rm a}\right)^{2}\left(\nu_{\rm a}/\nu_{\rm c}\right)^{1/3}L_{ \nu,\max}, & \nu<\nu_{\rm a} \\
\left(\nu/\nu_{\rm c}\right)^{1/3}L_{\nu,\max}, & \nu_{\rm a}<\nu<\nu_{\rm c} \\
\left(\nu/\nu_{\rm c}\right)^{-1/2}L_{\nu,\max}, & \nu_{\rm c}<\nu<\nu_{\rm s} \\
{\rm exp}(1-\left(\nu/\nu_{\rm s}\right))\left(\nu/\nu_{\rm s}\right)^{1/2}L_{ \nu,\max}, & \nu_{\rm s}<\nu,
\end{array}\right.
\end{array}
\end{eqnarray}	
for $\text { (a) } \nu_{\rm a}<\nu_{\rm c}<\nu_{\rm s}$, and~\begin{eqnarray}
\begin{array}{l}
L_{\nu}=\left\{\begin{array}{ll}
\left(\nu / \nu_{\rm c}\right)^{2}\left(\nu_{\rm c} / \nu_{\rm a}\right)^{3} L_{ \nu, \max }, & \nu<\nu_{\rm c} \\
\left(\nu / \nu_{\rm a}\right)^{5 / 2}\left(\nu_{\rm a} / \nu_{\rm c}\right)^{-1 / 2} L_{\nu, \max }, & \nu_{\rm c}<\nu<\nu_{\rm a} \\
\left(\nu / \nu_{\rm c}\right)^{-1 / 2} L_{\nu, \max }, & \nu_{\rm a}<\nu<\nu_{\rm s} \\
{\rm exp}(1-\left(\nu/\nu_{\rm s}\right))\left(\nu/\nu_{\rm s}\right)^{1/2}L_{\nu, \max }, & \nu_{\rm s}<\nu,
\end{array}\right.
\end{array}
\end{eqnarray}	
for $\text { (b) } \nu_{\rm c}<\nu_{\rm a}<\nu_{\rm s}$.
At later times, as~the ejecta expand and cool, the~electrons transition to the slow-cooling regime,  where
\begin{eqnarray}
\begin{array}{l}
L_{ \nu}=\left\{\begin{array}{ll}
\left(\nu/\nu_{\rm a}\right)^{2}\left(\nu_{\rm a}/\nu_{\rm s}\right)^{1/3}L_{\nu,\max}, & \nu<\nu_{\rm a} \\
\left(\nu/\nu_{\rm s}\right)^{1/3}L_{\nu,\max}, & \nu_{\rm a}<\nu<\nu_{\rm s} \\
{\rm exp}(1-\left(\nu/\nu_{\rm s}\right))\left(\nu/\nu_{\rm s}\right)^{1/2}L_{ \nu,\max}, & \nu_{\rm s}<\nu, \\
\end{array}\right.
\end{array}
\end{eqnarray}	
for $\text { (a) } \nu_{\rm a}<\nu_{\rm s}<\nu_{\rm c}$, and~\begin{eqnarray}
\begin{array}{l}
L_{ \nu}=\left\{\begin{array}{ll}
\left(\nu / \nu_{\rm syn}\right)^{2}\left(\nu_{\rm syn} / \nu_{\rm a}\right)^{(p+4)/2} L_{ \nu, \max }, & \nu<\nu_{\rm s} \\
{\rm exp}(1-\left(\nu/\nu_{\rm s}\right))\left(\nu/\nu_{\rm s}\right)^{1/2}L_{ \nu, \max }, & \nu_{\rm s}<\nu, \\
\end{array}\right.
\end{array}
\end{eqnarray}	
for $\text { (b) } \nu_{\rm s}<\nu_{\rm a}<\nu_{\rm c}$.
This spectral framework provides a robust description of thermal synchrotron emission from the forward shock, capturing its temporal and spectral evolution under various radiative~regimes.



\section{Results}

We present the theoretical multi-wavelength afterglow light curves of FRB 20171020A, considering
two different external environments: a uniform medium (ISM) and a wind-like medium. Our calculations are based on the thermal synchrotron radiation model described in Section~\ref{sec2}, and~the results are displayed in Figure~\ref{fig1} (uniform case) and Figure~\ref{fig:2} (wind case).
For each scenario, we compute light curves in the radio (1 GHz), optical (R-band), and~X-ray (1 keV) bands, and~compare them with the detection limits of Very Large Array (VLA), {which scales as $\propto t^{-1 / 2}$ for arbitrarily long exposure {times}.
{{In} survey mode, Legacy Survey of Space and Time (LSST) at the Vera C. Rubin Observatory reaches 24.5~mag in 30~s~\citep{2014ApJ...792L..21Y}}, and {the X-ray telescope (XRT) onboard the Neil Gehrels Swift Observatory  is $\propto t^{-1}$ early on and breaks to $\propto t^{-1 / 2}$ when $ F_{v}=2.0 \times 10^{-15}~\mathrm{erg} \mathrm{~cm}^{-2} \mathrm{~s}^{-1}$ at $t = 10^5$~s~\citep{Moretti2008MCAONP}}. 
We also consider both radiative (\(\epsilon = 1\)) and adiabatic (\(\epsilon = 0\)) cases, where \(\epsilon\) represents the efficiency of radiative losses of the blast wave. 
The kinetics of adiabatic evolution (\(\epsilon = 0\)) correspond to an optimistic scenario, whereas the kinetics of radiative evolution (\(\epsilon = 1\)) reflect a conservative~situation.

\begin{figure*}[htbp]
	\gridline{\fig{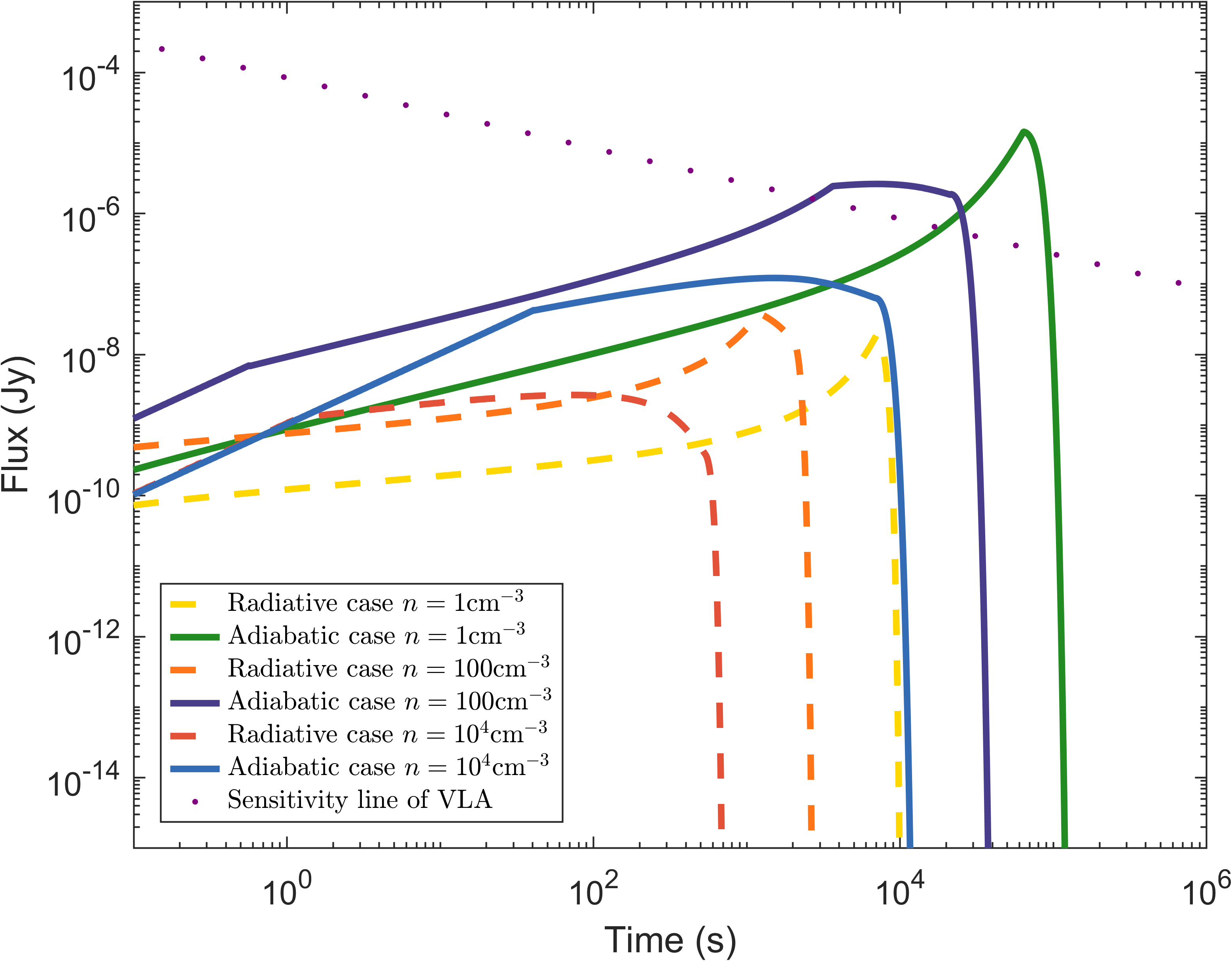}{0.3\textwidth}{(a)}
		\fig{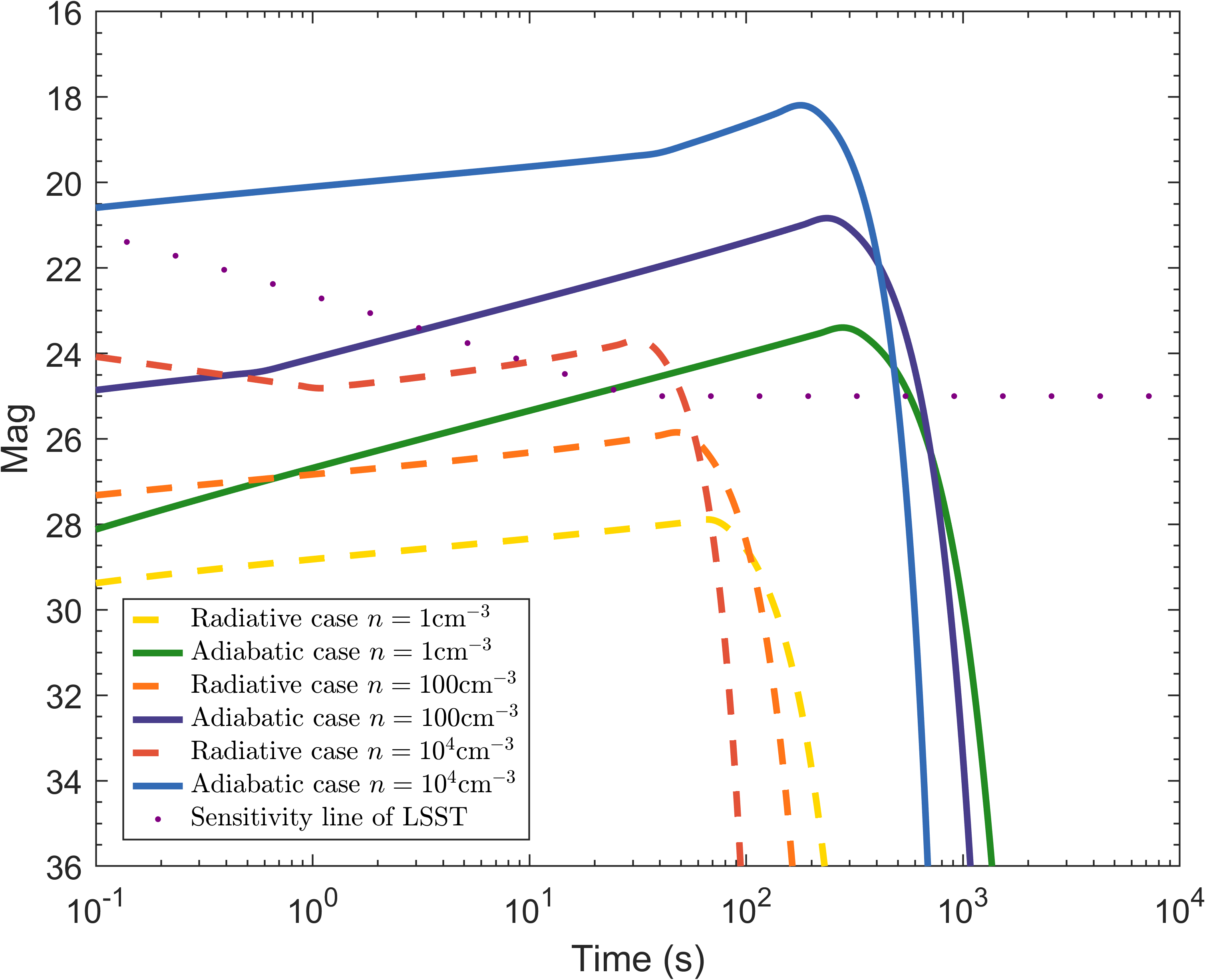}{0.29\textwidth}{(b)}
		\fig{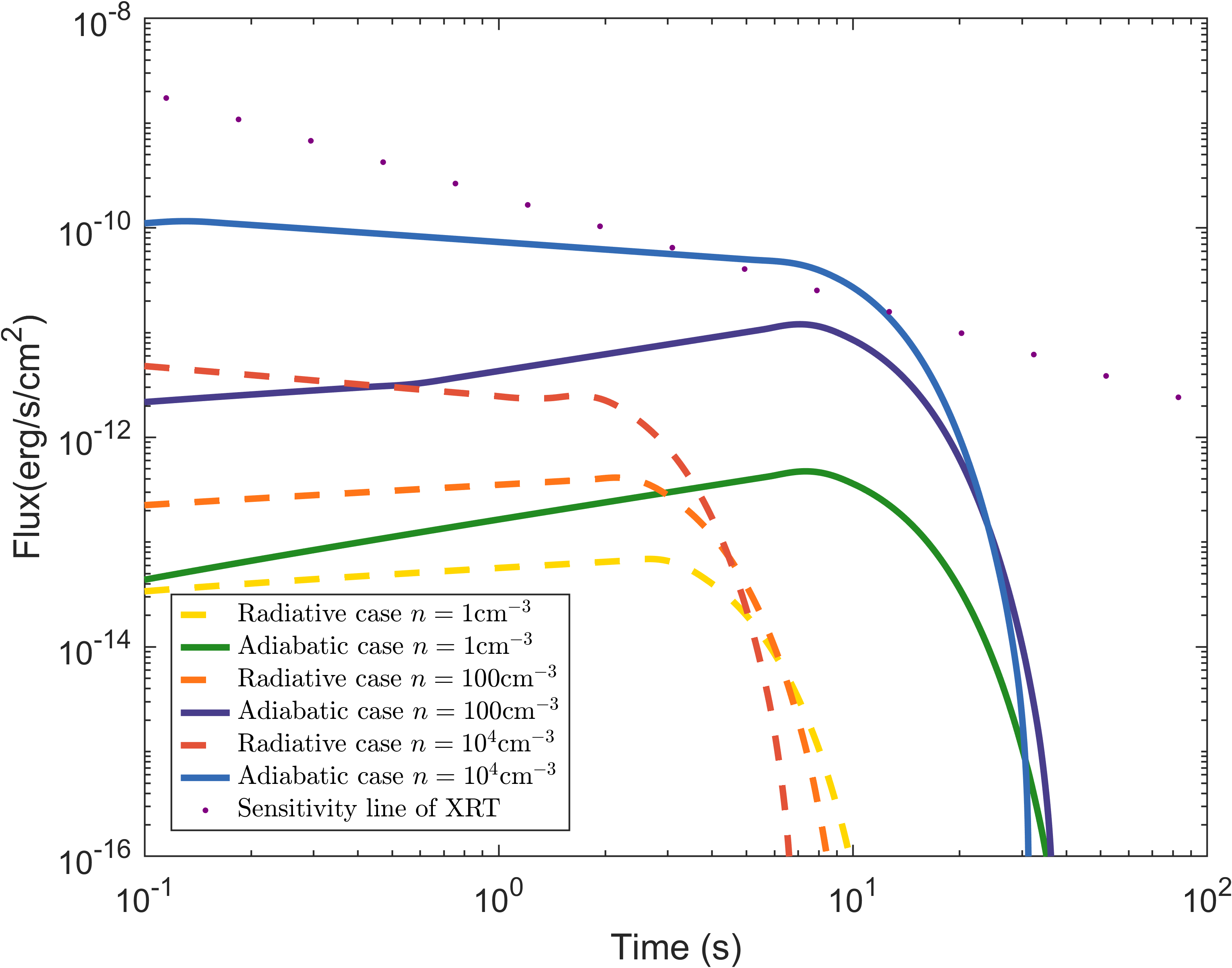}{0.3\textwidth}{(c)}
	}
	\caption{{Multi$-$wavelength} 
		afterglow light curves of FRB 20171020A in different uniform media. The~radio (1 GHz), optical (R-band), and~X-ray (1 keV) light curves are shown for both radiative (\(\epsilon = 1\), dashed lines) and adiabatic (\(\epsilon = 0\), solid lines) cases. The~detection limits of VLA (radio), LSST (optical), and~Swift/XRT (X-ray) are indicated by purple dotted~lines. Among them, (a), (b), (c) respectively represent the light variation curves under radio, optical and X-ray conditions. 
		\label{fig:1}}
\end{figure*}

\begin{figure*}[ht]
	\gridline{\fig{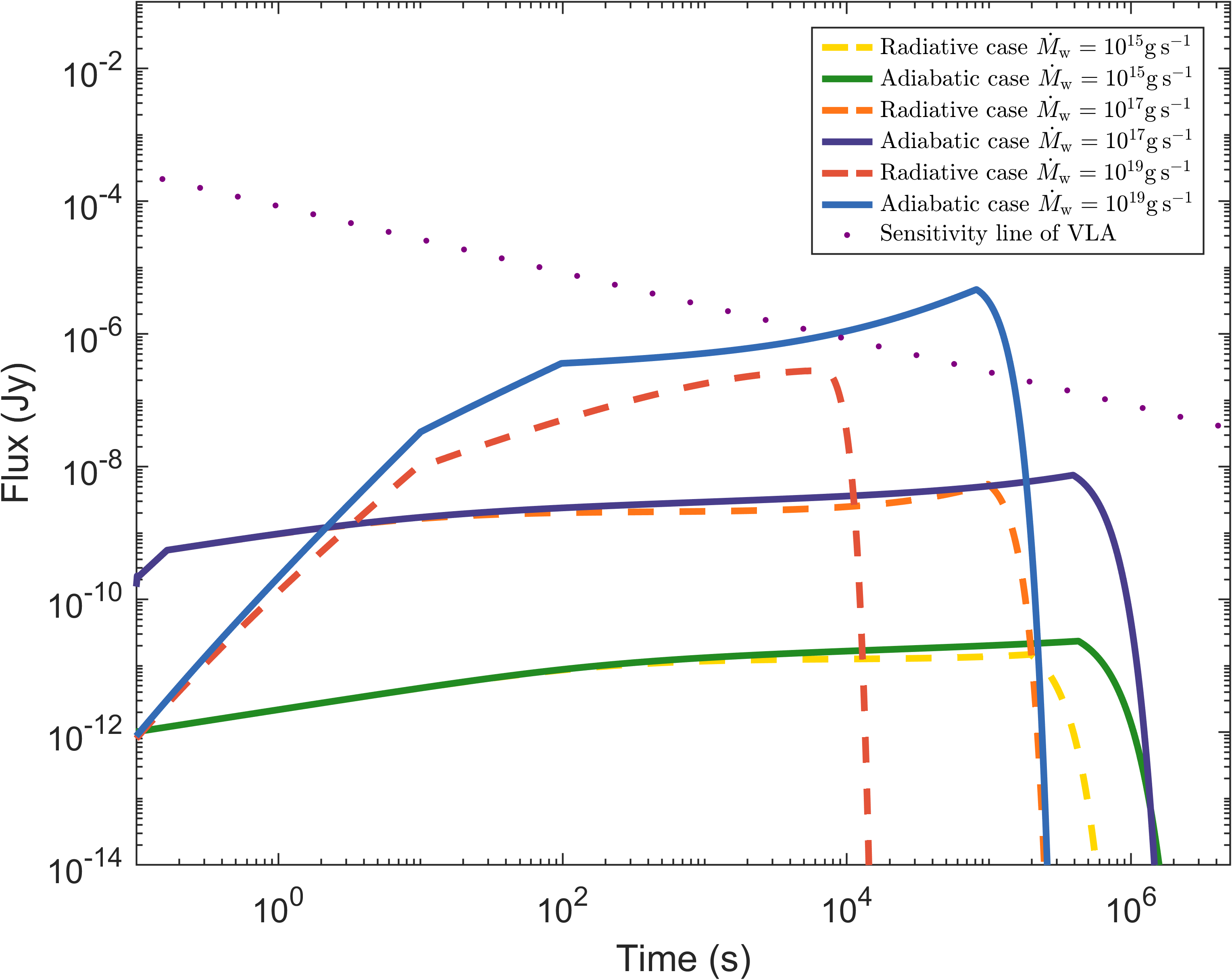}{0.3\textwidth}{(a)}
		\fig{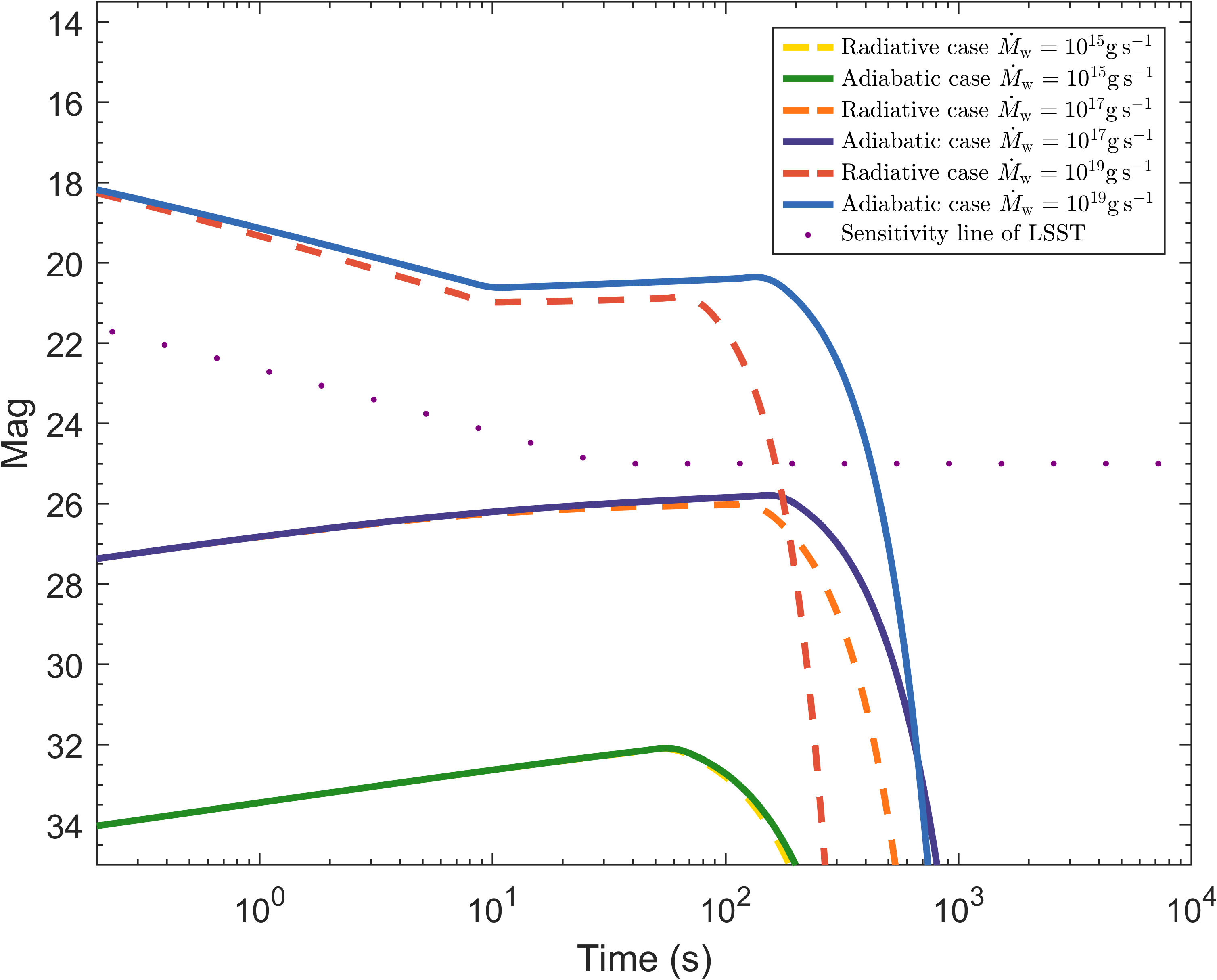}{0.29\textwidth}{(b)}
		\fig{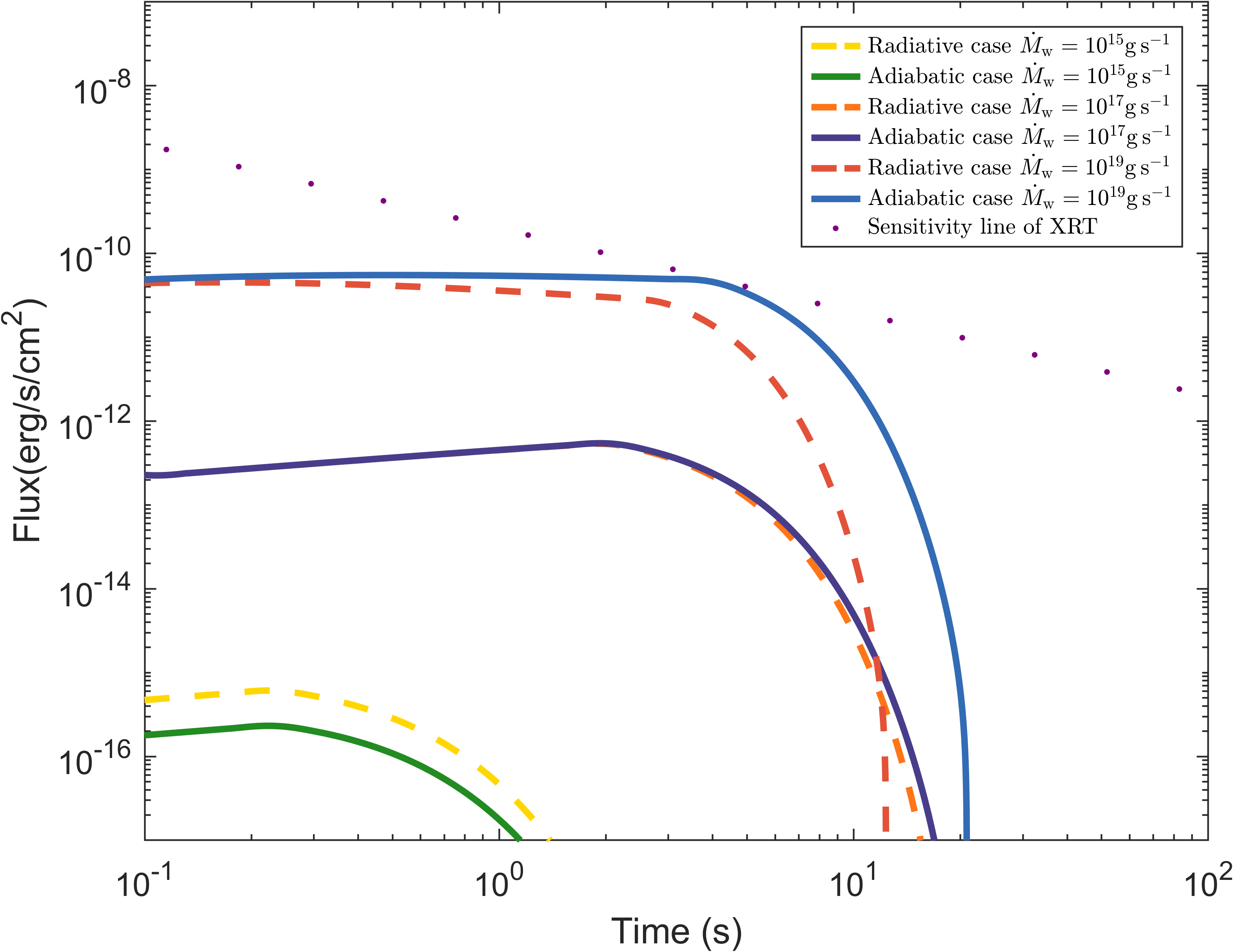}{0.3\textwidth}{(c)}		
	}
	\caption{Same as Figure~\ref{fig1}, but~for a wind-like environment. The~afterglow is generally brighter and longer-lasting {than the uniform medium case}, with~the radio and optical signals exceeding detection limits in the adiabatic case. However, the~X-ray afterglow remains undetectable with current instruments. Among them, (a), (b), (c) respectively represent the light variation curves under radio, optical and X-ray conditions.  
		\label{fig:2}}
\end{figure*}

\subsection{ Model Parameters and~Justification }
To compute the afterglow evolution of FRB 20171020A, we adopt key physical parameters based on theoretical models and observational constraints. The~chosen values for the kinetic energy of the ejecta (\(E_{\rm k}\)), initial Lorentz factor (\(\eta\)), and~external medium density (\(n_0\) for a uniform medium, \(\dot{M}_{\rm w}\) for a wind environment) are consistent with the magnetar model proposed by \citet{Metzger+2019}. Below, we detail the rationale behind each parameter~selection. 

The total kinetic energy of the relativistic ejecta is taken as  $E_k = 10^{43} \, \rm{erg}$.
This choice is motivated by comparisons with the Galactic FRB 200428D, which was associated with the magnetar SGR 1935+2154. The~radiation efficiency of FRB 200428D was estimated to be \(10^{-5}\) \citep{2020ApJ...899L..27M}, implying that the total energy released in the event is approximately \(10^5\) times the observed FRB energy. Given that FRB 20171020A has an observed isotropic energy of \(10^{38} \, \rm{erg}\)  \citep{2018ApJ...867L..12S}, we scale up the kinetic energy using the same efficiency factor, yielding \(E_k = 10^{43} \, \rm{erg}\). This is consistent with theoretical expectations for magnetar-powered relativistic outflows {\citep{Metzger+2019}}.

The initial Lorentz factor \(\eta\) governs the initial velocity of the ejecta and plays a crucial role in determining the afterglow evolution. We adopt different values for the uniform and wind medium scenarios. $\eta = 400$ for the uniform medium case, and~$\eta = 15$ for the wind medium case. These values are directly taken from the magnetar model of \citet{Metzger+2019}.  

The density of the surrounding medium affects how quickly the ejecta decelerate and determines the peak time and flux of the afterglow.  For~a uniform medium, we take the {number density $n$ in the range of \(1\)--\(10^4\,\mathrm{cm^{-3}}\) }following~\cite{2022MNRAS.517.5483C}.
This value is consistent with the conditions required for coherent maser emission in FRBs, as~derived in \citet{Metzger+2019}.  
For a stellar wind environment, the~density profile follows  
\[
n(r) = \frac{\dot{M}_w}{4\pi r^2 v_w m_p},
\]
where \( \dot{M}_w \) is the mass-loss rate, {and} $v_w$ is the speed of the wind. We adopt {$\dot{M}_w$ in the value range of $ 10^{15}$--$10^{19} \, \rm{g \, s^{-1}}$}, and~ $v_w=0.5c$  \citep{Metzger+2019,2022MNRAS.517.5483C}.
This is the characteristic mass-loss rate required in \citet{Metzger+2019} for magnetar-driven outflows interacting with a surrounding~wind.

These values ensure that our modeling remains consistent with FRB emission models while providing a robust framework for evaluating the detectability of FRB 20171020A’s afterglow.  
With these parameters established, the~following sections show how they influence the radio, optical, and~X-ray afterglows, and~how different environments impact the detectability of FRB 20171020A.
It is worth clarifying that the afterglow emissions discussed in this work refer to potential afterglows associated with future bursts from the source FRB 20171020A, rather than any afterglow following its previously detected burst. In~other words, our results provide predictions for the detectability of multi-wavelength afterglows assuming FRB 20171020A emits a new, bright burst in the future. Given its relatively close distance and high fluence, this source represents an ideal candidate for targeted follow-up observations. We emphasize that no afterglow has been observed from its earlier burst, and~our calculations are intended to guide expectations and strategies for future observational~campaigns.


\subsection{ Radio~Afterglow }
Figures~\ref{fig1}a and~\ref{fig:2}a display the radio afterglow light curves at 1 GHz in the uniform and wind-like environments, respectively. The~evolution of the radio afterglow shows distinct characteristics in these two cases. In~a uniform medium, Figure~\ref{fig1}a {shows the radio afterglow at 1 GHz. The~peak flux densities for different density values are \(1.59 \times 10^{-9}\, \rm{Jy}\) (\(n = 10^{4}\,\rm{cm}^{-3}\), red), \(8.60 \times 10^{-9}\, \rm{Jy}\) (\(n = 10^{2}\,\rm{cm}^{-3}\), orange), and~\(1.84 \times 10^{-8}\, \rm{Jy}\) (\(n = 1\,\rm{cm}^{-3}\), yellow) in the radiative case,}
and {\(2.67 \times 10^{-9}\, \rm{Jy}\) (\(n = 10^{4}\,\rm{cm}^{-3}\), blue), \(2.47 \times 10^{-6}\, \rm{Jy}\) (\(n = 10^{2}\,\rm{cm}^{-3}\), purple) and \(1.39 \times 10^{-5}\, \rm{Jy}\) (\(n = 1\,\rm{cm}^{-3}\), green) in the adiabatic case.}
However, these flux levels mostly remain below the detection sensitivity of VLA, implying that radio follow-up in this scenario would be challenging. In~a wind-like environment (Figure \ref{fig:2}a), the~radio afterglow is significantly brighter due to the denser medium. The~peak flux increases to  {\(2.34 \times 10^{-11}\, \rm{Jy}\)  ($\dot{M}_{\rm w} = 10^{15}\rm gs^{-1}$, green), \(7.45 \times 10^{-9}\, \rm{Jy}\)  ($\dot{M}_{\rm w} = 10^{17}\rm gs^{-1}$, purple) and \(4.65 \times 10^{-6}\, \rm{Jy}\)  ($\dot{M}_{\rm w} = 10^{19}\rm gs^{-1}$, blue) in the adiabatic case, and~\(1.45 \times 10^{-11}\, \rm{Jy}\)  ($\dot{M}_{\rm w} = 10^{15}~\rm gs^{-1}$, yellow), \(5.30 \times 10^{-9}\, \rm{Jy}\)  ($\dot{M}_{\rm w} = 10^{17}~\rm gs^{-1}$, orange) and \(2.70 \times 10^{-7}\, \rm{Jy}\)  ($\dot{M}_{\rm w} = 10^{19}~\rm gs^{-1}$, red) in the radiative case. }
Importantly, the~radio flux for {the} adiabatic case exceeds the VLA sensitivity, suggesting that a wind environment provides a better chance for radio afterglow detection.  These results indicate that the detectability of the radio afterglow strongly depends on the external medium density. If~FRB 20171020A originated in a wind-like environment, {radio observations at {the} 
	$\mu \rm{Jy}$ level with facilities like the VLA or SKA are capable of detecting such afterglows.} 

\subsection{Optical~Afterglow}
Figures~\ref{fig1}b and \ref{fig:2}b show the optical afterglow in the R-band, compared to the LSST sensitivity limit. 
In a uniform medium (Figure \ref{fig1}b), the~peak magnitudes are {27.9 (\mbox{\(n = 1\,\rm{cm}^{-3}\),} yellow), 25.8 (\(n = 10^{2}\,\rm{cm}^{-3}\), orange) and 23.6 (\(n = 10^{4}\,\rm{cm}^{-3}\), red) in the radiative case. and 23.4 (\(n = 1\,\rm{cm}^{-3}\), green), 20.8 (\(n = 10^{2}\,\rm{cm}^{-3}\), purple) and 18.1 (\mbox{\(n = 10^{4}\,\rm{cm}^{-3}\)}, blue) in the adiabatic case.}
In a wind-like environment (Figure \ref{fig:2}b), the~peak magnitudes improve to {32.0 (\(\dot{M}_{\rm w} = 10^{15}\,\rm{g\,s^{-1}}\), yellow), 26.0 (\(\dot{M}_{\rm w} = 10^{17}\,\rm{g\,s^{-1}}\), orange) and 20.8 (\(\dot{M}_{\rm w} = 10^{19}\,\rm{g\,s^{-1}}\), red) in the radiative case. and 32.0 (\(\dot{M}_{\rm w} = 10^{15}\,\rm{g\,s^{-1}}\), green), 25.7 (\(\dot{M}_{\rm w} = 10^{17}\,\rm{g\,s^{-1}}\), purple) and 20.3 (\(\dot{M}_{\rm w} = 10^{19}\,\rm{g\,s^{-1}}\), blue) in the adiabatic case.}
These results suggest that LSST has an excellent chance of detecting the optical afterglow, but~only if observations are conducted within a few hundred seconds post-burst. Rapid optical follow-up is crucial to capture the optical~afterglow.

\subsection{X-Ray~Afterglow}
Figures~\ref{fig1}c and \ref{fig:2}c show the X-ray afterglow at 1 keV, compared to the Swift/XRT sensitivity.  
In a uniform medium (Figure \ref{fig1}c), the~X-ray afterglow is too faint to be detected.  
In a wind-like environment (Figure \ref{fig:2}c), the~X-ray flux is slightly higher but still remains below the Swift/XRT sensitivity threshold, even for the most optimal scenario.  
These results indicate that current X-ray facilities are unlikely to detect the afterglow of FRB 20171020A. However, future wide-field X-ray telescopes may improve detection~prospects.

\subsection{Detectability Across Parameter~Space}

{
	To more comprehensively explore the detectability of afterglows across a wide parameter space, we present contour plots of the peak R-band magnitude as a function of kinetic energy and environmental density simultaneously, for~both a uniform medium and a wind-like environment. Given our results that the optical band typically offers the highest detectability, we focus on the optical afterglow to assess observational prospects.
}

{In our modeling, we consider two limiting cases for the shock evolution: adiabatic and radiative. In~the adiabatic case, radiative energy losses are negligible, and most of the shock-generated internal energy is retained in the outflow. This corresponds to a radiation efficiency close to 0. In~contrast, the~radiative case assumes that nearly all of the internal energy is promptly radiated away, i.e.,~with 100\% efficiency. These two cases represent the optimistic (adiabatic) and conservative (radiative) limits for predicting afterglow brightness. By~analyzing both scenarios, we can bracket the expected range of optical flux and provide a robust assessment of detectability under varying physical conditions.
}

{Figure~\ref{fig:3} displays the contours in the \(E_K\)–\(n\) plane for a uniform-density environment, where \(E_K\) ranges from \(10^{40}\) to \(10^{43}\,\mathrm{erg}\) and the number density spans \(1\)–\(10^4\,\mathrm{cm^{-3}}\). The~results show that the detectability of the optical afterglow varies significantly between the adiabatic and radiative regimes. Notably, in~the adiabatic case, detectable afterglows are possible for kinetic energies  $>10^{41}$ erg, while in the radiative case, the~threshold for detection shifts to higher energies {$ >10^{43}$ erg }due to enhanced energy losses.
	Figure~\ref{fig:4} shows similar contours in the \(E_K\)–\(\dot{M}_{\rm w}\) plane for a wind-like environment. As~in Figure~\ref{fig:3}, the~black dashed lines indicate the detection threshold of the LSST in the optical band. The~contours reveal that only when the circum-burst environment is sufficiently dense (i.e., large \(\dot{M}_{\rm w}\)) can a bright enough optical afterglow be produced to exceed the LSST detection limit.
}

{These results suggest that the detectability of the afterglow from FRB 20171020A depends sensitively on both the energetics of the outflow and the density of the surrounding medium. In~particular, detectable optical afterglows are expected only for events with kinetic energy \(\gtrsim 10^{43}\,\mathrm{erg}\) and dense local environments. 
}

\begin{figure*}[htbp]
	\gridline{\fig{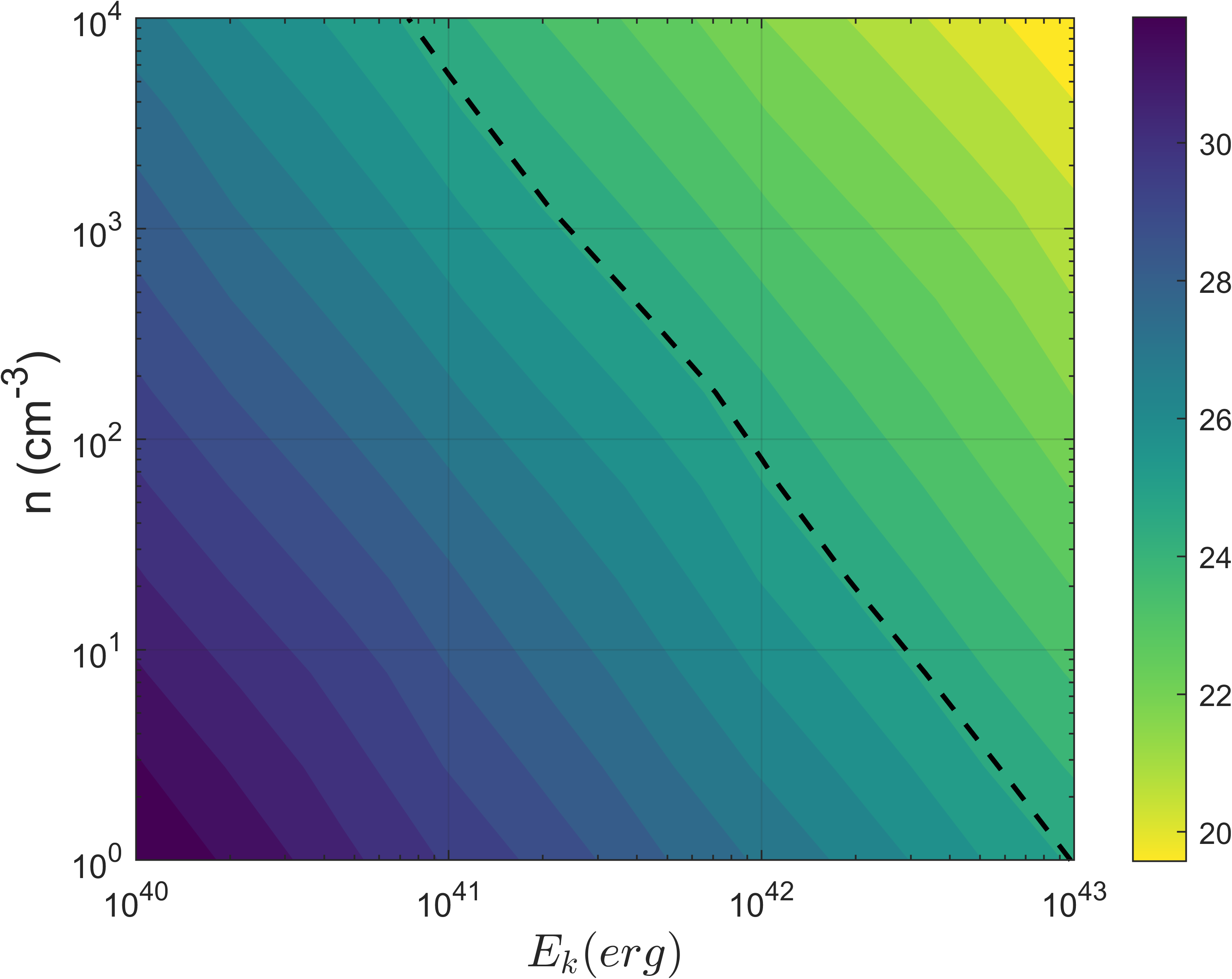}{0.3\textwidth}{(a)}
		\fig{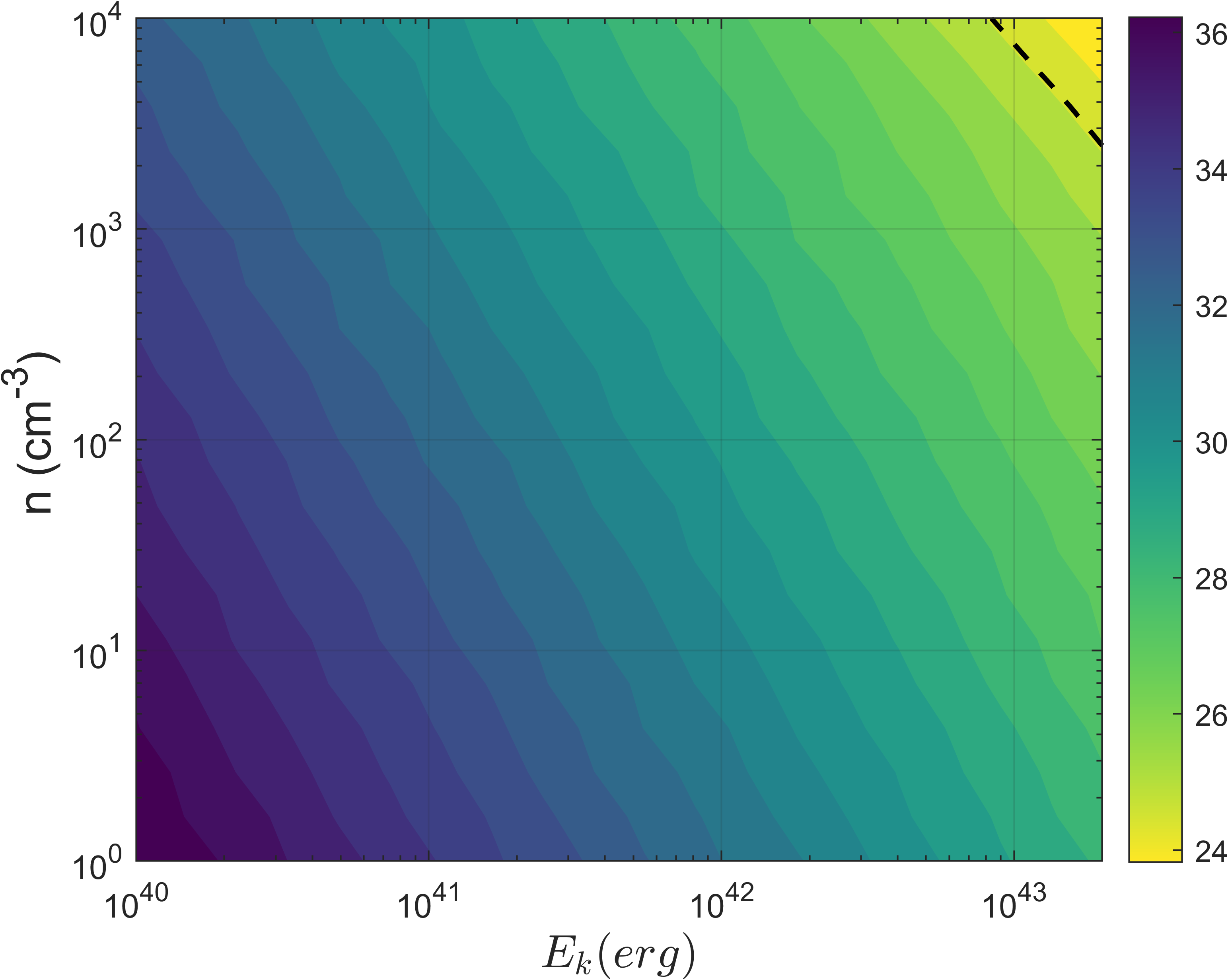}{0.29\textwidth}{(b)}
	}
	\caption{Contour of optical peak flux in the \(E_K\)$-$\(n\) plane in a uniform medium. Different colors correspond to different optical peak flux levels (magnitude in the R-band). The~black dashed lines indicate the LSST observational upper limits. Panel (\textbf{a}) represents the adiabatic case, while Panel (\textbf{b}) corresponds to the radiative case.
		\label{fig:3}}
\end{figure*}

\begin{figure*}[htbp]
	\gridline{\fig{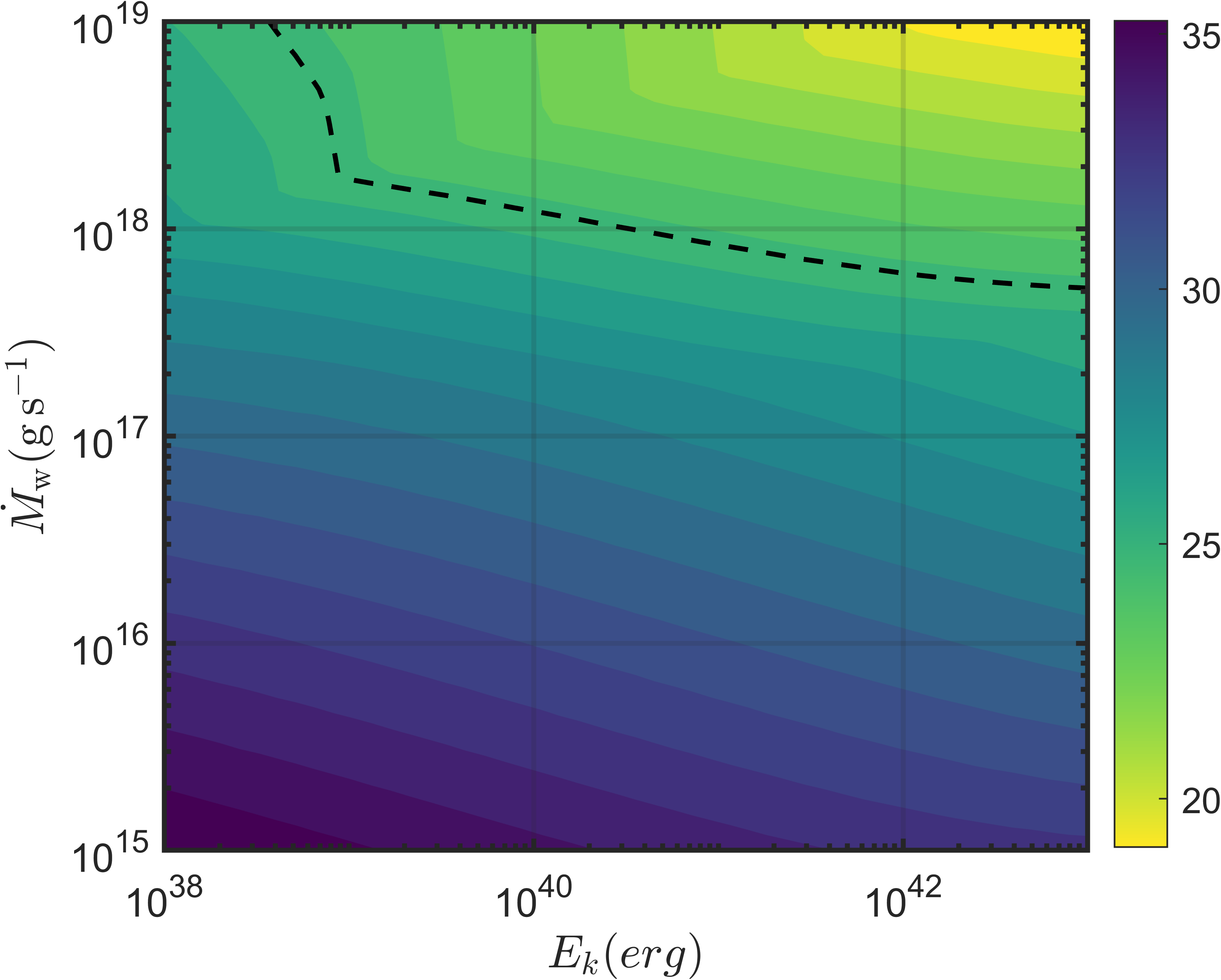}{0.3\textwidth}{(a)}
		\fig{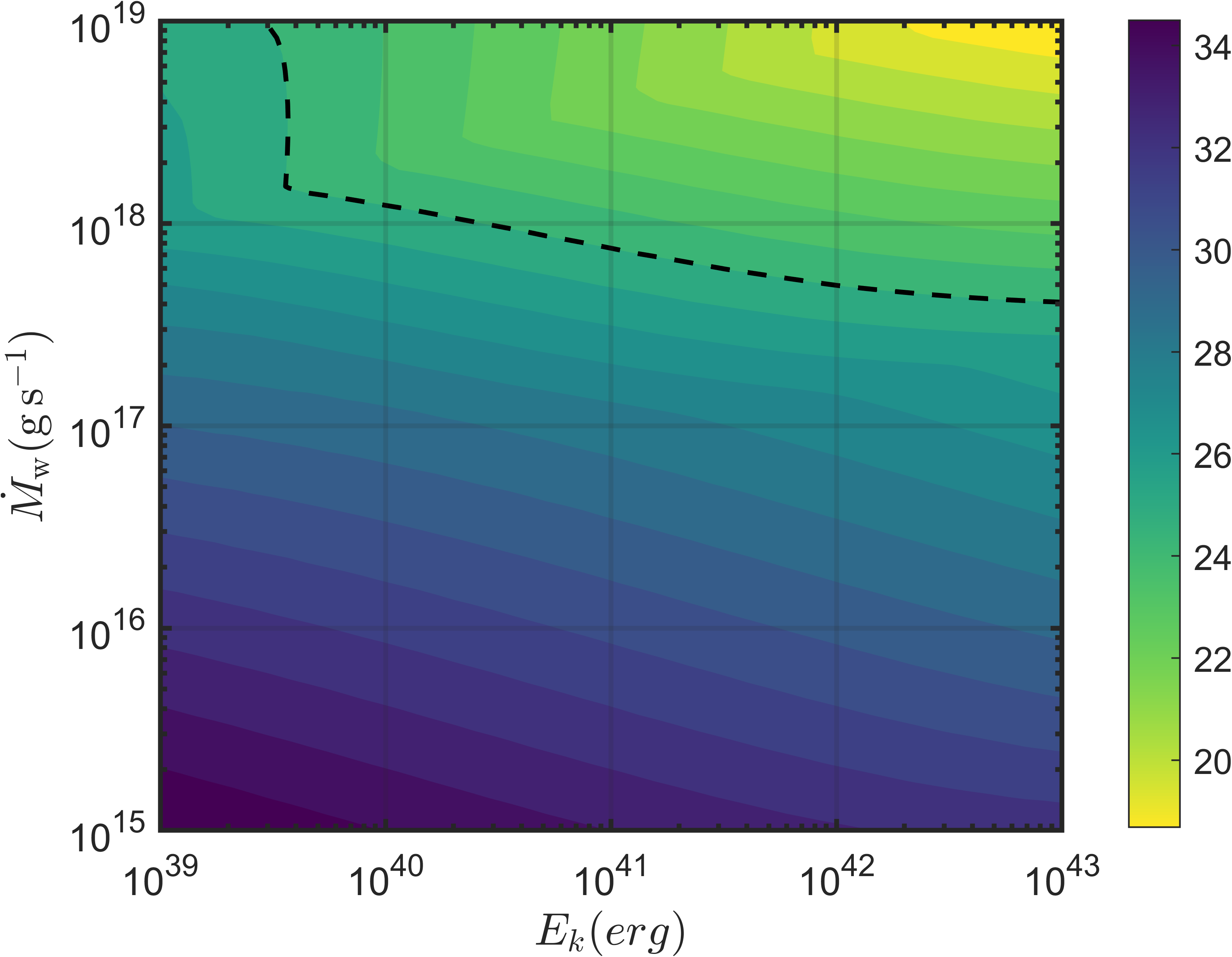}{0.29\textwidth}{(b)}
	}
	\caption {{Contour}  of optical peak flux in the \(E_K\)$-$$\dot{M}_{\rm w}$ plane for a wind-like environment. Different colors correspond to different optical peak flux levels (magnitude in the R-band). The~black dashed lines indicate the LSST observational upper limits. Panel ({a}) represents the adiabatic case, while \mbox{Panel ({b})} corresponds to the radiative case.
		\label{fig:4}}
\end{figure*}

\section{Conclusions and~Discussions}

{We} have investigated the possible multi-wavelength afterglow of FRB 20171020A, one of the closest and brightest extragalactic FRBs, using the thermal synchrotron emission model in a highly magnetized shock scenario. By~considering two different external environments---a uniform medium and a stellar wind environment—we examined the afterglow evolution in radio, optical, and~X-ray bands and assessed its detectability with current and future observational~facilities.

Our results indicate that the optical afterglow is the most promising for detection, as~it reaches a brightness significantly exceeding the LSST detection threshold. In~both environmental scenarios, the~optical afterglow is well above LSST’s sensitivity, making it a strong candidate for follow-up observations. However, due to its rapid fading, it requires early observations within a few hundred seconds post-burst to maximize the likelihood of detection. This highlights the importance of rapid-response optical surveys in identifying FRB~afterglows.

The radio afterglow, on~the other hand, is much less certain. Only in the most favorable conditions, such as in a dense stellar wind environment, does the predicted radio flux exceed the VLA sensitivity threshold. Even in this case, detection remains challenging, and~deep radio follow-up is required. In~a uniform medium, the~radio flux remains too low for current radio telescopes to detect. Therefore, while radio observations remain an important tool for constraining the FRB environment, their success depends strongly on the external medium~properties.

The X-ray afterglow is expected to be too faint for detection with {Neil Gehrels Swift/XRT}, regardless of the environment. This suggests that current X-ray facilities are unlikely to observe the high-energy counterpart of FRB 20171020A. However, future wide-field sensitive X-ray observatories  may improve detection~prospects.

Overall, our results demonstrate that the circum-burst environment plays a crucial role in shaping the afterglow properties. If~FRB 20171020A is surrounded by a dense stellar wind, its afterglow is brighter and longer-lasting, improving the chances of detection. If~it is instead propagating through a uniform medium, its afterglow is significantly weaker, making detection more difficult, especially in the radio~band.

Given these findings, the~best strategy for detecting FRB afterglows involves rapid optical follow-up with observational facilities like LSST, which offers the highest probability of success. Deep radio follow-up with VLA or SKA is still valuable but will likely only yield detections in favorable environmental conditions. Meanwhile, future X-ray surveys may help place additional constraints on FRB afterglow properties.
As FRB discoveries continue to grow, nearby and bright events like FRB 20171020A provide an excellent opportunity to test theoretical models and refine our understanding of FRB progenitors, their energy release mechanisms, and~their connection to magnetars. Multi-wavelength follow-up of similar FRBs will be crucial for revealing the true nature of these  mysterious radio~transients.

{Furthermore, it is natural to ask how frequently such nearby, energetic FRBs—capable of producing detectable afterglows—might occur and be observed in practice. Although~the focus of this work is on FRB 20171020A and the possible multi-wavelength afterglow that may accompany its future bright bursts, we believe that discussing the general occurrence rate and detectability of nearby FRBs similar to FRB 20171020A is interesting. Such analysis also provides broader context and motivates the search for other nearby FRBs with similar observational potential.}

{From our model calculations, which explore a broad parameter space including variations in ambient density and dynamics, we find that afterglows become potentially detectable only when the kinetic energy of the ejector exceeds $\sim$10$^{43}$ erg. This threshold holds across various plausible environments (e.g., uniform or wind-like media) and dynamics.
	Given that typical FRB radiation efficiencies have {been} adopted to be $\sim$10$^{-5}$ in this paper, we infer \(E_{\mathrm{FRB}} > 10^{38}\) erg as a practical threshold to identify nearby FRBs whose afterglows are realistically detectable.}

Based on current estimates of the volumetric rate of FRBs with \(E_{\mathrm{FRB}} > 10^{38}\) erg, the~occurrence rate is approximately $\sim$10$^5$ Gpc\(^{-3}\) yr\(^{-1}\) \citep{2023ApJ...944..105S}. Integrating over a volume up to 100 Mpc yields an all-sky rate of about 100 such nearby events per year. Considering that CHIME has a field of view covering approximately 0.3\% of the sky, we estimate that it may detect $\sim$0.3 nearby FRBs per year with sufficient energy to produce detectable afterglows. This low but non-negligible rate underscores the importance of rapid, multi-wavelength follow-up for exceptionally bright and nearby FRBs such as FRB 20171020A, and~reinforces the need to identify and monitor nearby high-energy FRBs as prime candidates for afterglow~searches.

Although this rate is low, it is not prohibitive. The greater challenge arises from the short lived nature of the predicted optical afterglow, which typically peaks and fades within $\sim$100 s of the FRB event demanding extremely prompt follow-up observations. Assuming a typical 10\% duty cycle for a single optical telescope, the~probability of catching such an event is correspondingly small, implying that it could take approximately 30 years of continuous operation to detect a single l{afterglow} 

It is important to emphasize that our estimate represents a conservative lower bound. The~assumed parameters—such as the minimum energy threshold and ambient environments—were chosen to reflect the most restrictive conditions under which the optical afterglow could still be marginally detectable. The~actual detection probability may be substantially higher under more favorable (but still realistic) physical conditions. However, a~reliable quantification of this would require a detailed Monte Carlo simulation across a high-dimensional parameter space, which lies beyond the scope of this work. This clearly illustrates the practical difficulty of detecting these rare but valuable transients, even for bright and nearby sources like FRB~20171020A.

Despite these challenges, several practical strategies could significantly improve the likelihood of detecting such afterglows. These include the deployment of robotic optical telescopes at geographically distributed sites to increase temporal and spatial coverage; the implementation of low-latency FRB detection pipelines capable of rapid DM-based distance estimation and event prioritization; and expanding the field of view and sensitivity of current and next-generation radio facilities to better capture rare nearby events in real~time.

\section*{Acknowledgments}
This work is supported by the National  Natural Science Foundation of China (grant No. 12203013) and the Guangxi Science Foundation (grant Nos.AD22035171  and 2023GXNSFBA026030).

\bibliography{ref}

\end{document}